\documentclass[11pt]{article}
\usepackage{moriond,epsfig}

\newcommand{\gtsimeq}{\raisebox{-0.6ex}{$\,\stackrel 
        {\raisebox{-.2ex}{$\textstyle >$}}{\sim}\,$}}

\begin{document}
\vspace*{4cm}
\title{The ELAIS Deep X-ray Survey}

\author{C.J.\,Willott}

\address{Astrophysics, University of Oxford, Keble Road, Oxford OX1
3RH, U.K.}

\author{O.\,Almaini, J.\,Manners, O.\,Johnson, A.\,Lawrence, J.S.\,Dunlop, R.G.\,Mann}

\address{Institute for Astronomy, University of Edinburgh, Blackford
Hill, Edinburgh EH9 3HJ, U.K.}

\author{I.\,Perez-Fournon, E.\,Gonzalez-Solares, F.\,Cabrera-Guerra}

\address{Instituto de Astrof\'\i sica de Canarias, C/ Via Lactea s/n,
38200 La Laguna, Tenerife, Spain}

\author{S.\,Serjeant}

\address{Unit for Space Sciences and Astrophysics, The University of
Kent, Canterbury, Kent CT2 7NR, UK}

\author{S.J.\,Oliver}

\address{Astronomy Centre, CPES, University of Sussex, Falmer,
Brighton, Sussex BN1 9QJ, UK}

\author{M.\,Rowan-Robinson}

\address{Astrophysics Group, Blackett Laboratory, Imperial College of
Science Technology \& Medicine (ICSTM), Prince Consort Rd, London SW7
2BZ, UK}

\maketitle

\abstracts{We present initial follow-up results of the ELAIS Deep
X-ray Survey which is being undertaken with the Chandra and XMM-Newton
Observatories. 235 X-ray sources are detected in our two 75 ks ACIS-I
observations in the well-studied ELAIS N1 and N2 areas. 90\% of the
X-ray sources are identified optically to $R=26$ with a median
magnitude of $R=24$. We show that objects which are unresolved
optically (i.e. quasars) follow a correlation between their optical
and X-ray fluxes, whereas galaxies do not. We also find that the
quasars with fainter optical counterparts have harder X-ray spectra,
consistent with absorption at both wavebands. Initial spectroscopic
follow-up has revealed a large fraction of high-luminosity Type 2
quasars. The prospects for studying the evolution of the host galaxies
of X-ray selected Type 2 AGN are considered.}
 
\section{Introduction}

X-ray astronomy is moving into a new era with the successful launch of
the Chandra and XMM-Newton observatories. The combination of large
collecting area, high spatial resolution and wide spectral range
enable new surveys to probe deeper than ever before and resolve most
of the hard X-ray (2-10 keV) background (XRB). The peak intensity of
the XRB lies at $30$ keV and models of the XRB spectrum (e.g. Comastri
et al. 1995; Gilli et al. 1999) predict that obscured AGN with very
hard spectra must be much more common in the Universe than naked
quasars. Ordinary, optically selected AGN perhaps account for only
10-20\% of the total hard XRB. The direct implication is that most of
the accretion activity in the Universe is absorbed.

We are conducting a deep X-ray survey with Chandra and XMM-Newton in
fields which have been well-studied at other wavebands. In each of the
two fields, designated N1 and N2, Chandra ACIS-I observations of
duration 75 ks have been made. A total of 235 X-ray sources are
detected in the full band images above a flux limit of $S_{0.5-10{\rm
keV}}>1.3 \times 10^{-15}~ {\rm erg ~cm}^{-2}~{\rm s}^{-1}$. We have
also been awarded time for a 150 ks observation of N2 with
XMM-Newton. Multi-wavelength coverage is very important for
understanding the nature of the X-ray sources. Our fields lie within
the European Large-Area ISO Survey (ELAIS) and have been observed with
ISO at 7, 15, 90 and 175 $\mu$m (Oliver et al. 2000) and with the VLA
at 1.4 GHz (Ciliegi et al. 1999). One of the two X-ray fields is
co-incident with the widest survey yet made with SCUBA on the JCMT --
the UK Submillimetre Consortium's 8 mJy Survey (Dunlop et al. 2000).

The goals of our survey are to obtain a better understanding of the
nature of the AGN responsible for the X-ray background (or
equivalently, the accretion history of the Universe) and the host
galaxies they reside in. Some of the questions to be answered are:
What is the connection between AGN activity and star-formation? Do
X-ray AGN live in old, evolved galaxies or young galaxies with high
levels of star-formation? What are the clustering properties of X-ray
sources, both with other X-ray sources and with the general galaxy
population?  How much of the accretion energy of the Universe is
obscured and reprocessed by dust? Is the dust distribution
well-represented by torus models with typical galactic gas-to-dust
ratios? What X-ray luminosity dependence is there on host galaxy/torus
properties? The answers to these questions will come from detailed
follow-up of deep X-ray surveys. 

Note that the small sky areas covered in each pointing with Chandra
(equivalent to $\sim 7\times 7$ Mpc at $z \gtsimeq 1$) mean that the
effects of small-scale clustering and cosmic variance are
important. For example, in the N1 field the density of X-ray sources
brighter than a given flux-limit is approximately 30\% higher than
that in N2. Thus, no single ultradeep survey (such as the Chandra Deep
Field South or the Hubble Deep Field North) will be able to accurately
determine the contribution to the XRB, the redshift distribution or
luminosity function of X-ray AGN. Only by combining the results of
several deep surveys will such results be obtainable.

A flat cosmology with parameters $H_0=70~ {\rm km~s^{-1}Mpc^{-1}}$,
$\Omega_{\mathrm M}=0.3$ and $\Omega_\Lambda=0.7$ is assumed throughout.

\section{Optical Identification of the ELAIS Deep X-ray Survey}

\begin{figure} 
\vspace{-0.7cm} 
\epsfxsize=1.00\textwidth
\epsfbox{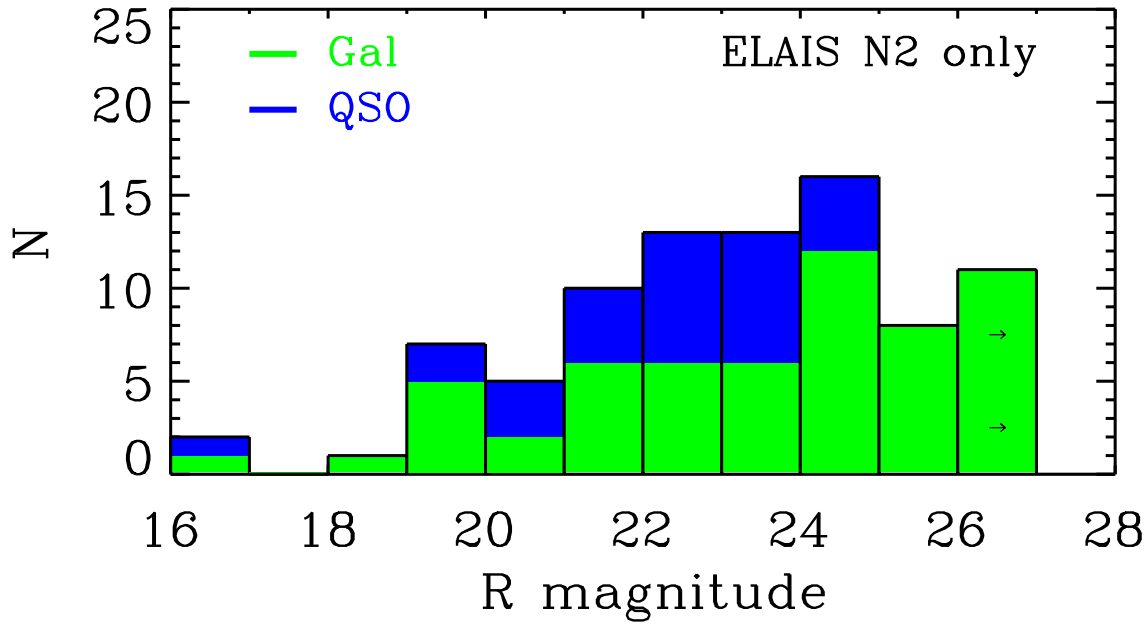} 
{\caption {Magnitude histogram of Chandra source optical
identifications in the ELAIS N2 region. The histogram is split into
those objects with a quasar-like unresolved optical appearance and
those which appear to be galaxies.}}  
\end{figure}

Optical imaging data covering the Chandra fields has been obtained
with the William Herschel Telescope (WHT) Prime-Focus Camera and the
Isaac Newton Telescope (INT) Wide-Field Camera. The WHT data are only
in N2 and reach to limiting Vega magnitudes of $R\approx 26$ and
$I\approx 26$. INT data in both fields reach limiting AB magnitudes of
$g'\approx 25.5$, $r'\approx 25$, $i'\approx 24.5$. In the
near-infrared there is full coverage of both fields at $H$-band with
CIRSI on the INT and partial $K$-band coverage from INGRID on the WHT
and UFTI on the UKIRT. Full details of all these observations will be
presented in forthcoming papers. 

For initial identification of the X-ray sources detected in the full
(0.5-10 keV) band, we consider only the WHT $R$-band imaging in N2 and
the INT $r'$-band imaging in N1. The X-ray and optical astrometric
frames are registered using $\approx 20$ X-ray sources which are
optically bright. The adopted identification procedure takes into
account the positional uncertainty of the Chandra source, which is
dependent upon the number of photons detected and the off-axis angle
(since the PSF is degraded as one moves further from the centre of the
field). The positional uncertainties range from 0.5 to 2.5 arcsec. A
modified version of the likelihood ratio procedure (Sutherland \&
Saunders 1992) was used to determine the likelihood of possible
identifications. Full details of the adopted method will be given in
Gonzalez-Solares et al. (in prep.). The median magnitude of the X-ray
source counterparts is $R\approx 24$ with 90\% identified to a limit
of $R=26$. Our results show a small, but non-negligible, percentage of
chance associations (5\%) of optical counterparts to Chandra sources,
most of which would occur near the field edges where the Chandra PSF
is quite large. Note that this will be a serious problem for deep
XMM-Newton surveys which have lower spatial resolution than
Chandra. Combining XMM-Newton and Chandra data will help to alleviate
this problem.

The good seeing (0.8 arcsec) of the optical imaging allows us to make
a morphological classification of the optical counterparts of X-ray
sources. We have used the Sextractor program to generate image source
catalogues and the parameter CLASS\_STAR is used to determine whether
an object has an unresolved (stellar) or resolved (galaxy)
morphology. The unresolved sources correspond predominantly to quasars
in which the light from the active nucleus substantially outshines
that of its host galaxy, whereas the resolved counterparts will be
dominated by starlight with any AGN component absorbed or
intrinsically weak. In Figure 1 we show a histogram of the magnitudes
of X-ray source identifications in the N2 region, for which the
star-galaxy separation works well down to $R=25$. There are
approximately equal numbers of quasars and galaxies at magnitudes
brighter than $R=24$ and then galaxies begin to dominate at fainter
magnitudes. We note that our results appear inconsistent with those of
Giacconi et al. (2001) who quote that two-thirds of the counterparts
to Chandra sources in the Chandra Deep Field South (CDFS) appear
point-like. However, high-resolution studies with the HST of a subset
of the CDFS sources (Schreier et al. 2001), shows an unresolved
fraction of about one third, similar to that in Fig.1.

\begin{figure}[t] 
\vspace{-0.5cm} 
\epsfxsize=1.05\textwidth
\epsfbox{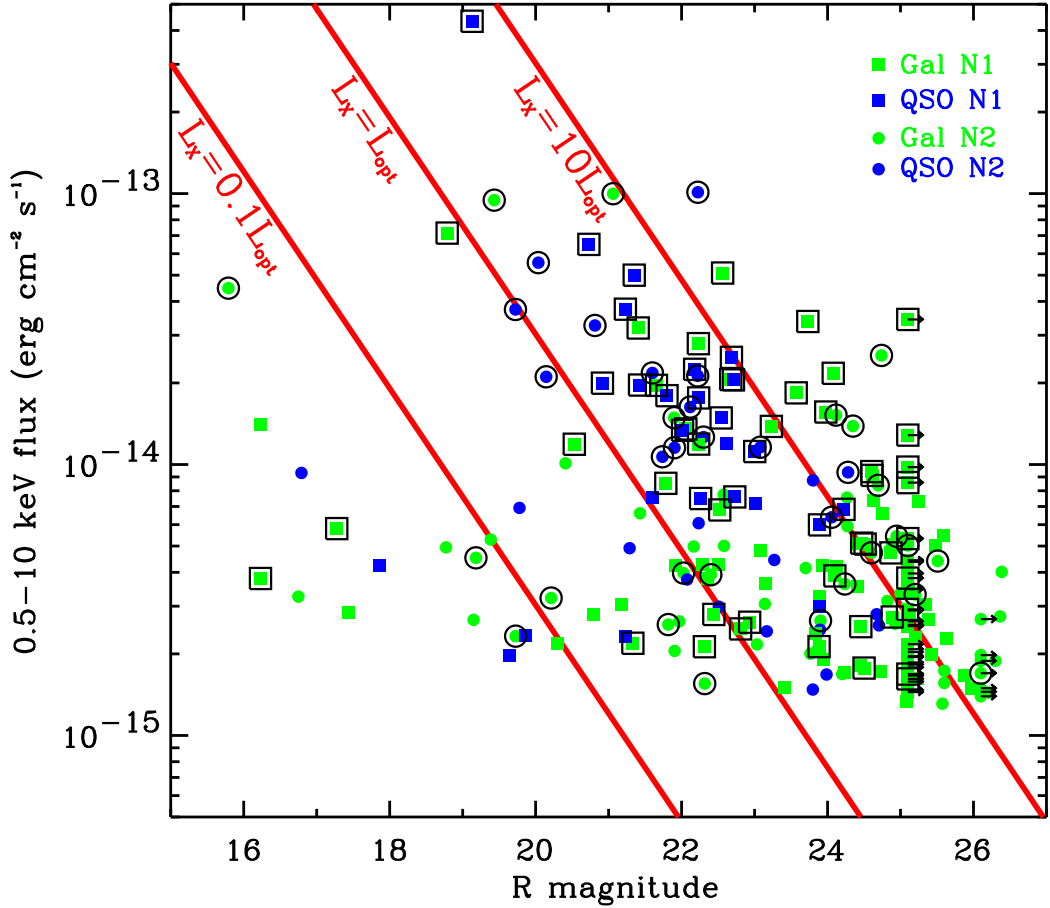} 

{\caption {0.5-10 keV flux against $R$ magnitude for ELAIS Chandra
sources. Sources also detected in the hard band images (2-10 keV) are
shown as symbols with a black border.}}

\end{figure}

Figure 2 plots the X-ray flux of sources against their optical
magnitudes. The quasars clearly show a well-defined correlation with
most sources lying in the band $1< L_{\rm X}/L_{\rm opt} < 10$. This
correlation is similar to that found for soft X-ray selected quasars
(Hasinger et al. 1999). We note that at least one of the few `quasar'
objects to the left of this diagram is spectroscopically confirmed as
a star. In contrast to the quasars, the distribution of galaxies on this plot
is much less well-defined. This is because for the galaxies, the
optical light is dominated by the stellar population, which is not
tightly correlated with the X-ray flux coming, presumably, from the
nucleus. There are many galaxies with $L_{\rm X}/L_{\rm opt} > 10$,
which are presumably Type-2 quasars in which the optical flux is much
more readily depressed by absorption than the hard X-rays.  The
correlation between optical and X-ray flux exhibited by the quasars,
but not the galaxies, explains why the fraction of galaxy counterparts
increases fainter than $R=24$, since it is at this magnitude where
$L_{\rm X} \approx L_{\rm opt}$ at our X-ray flux limit.

\section{Comparison of optical properties with X-ray hardness}

\begin{figure}[t] 
\vspace{-0.5cm} 
\epsfxsize=1.05\textwidth
\epsfbox{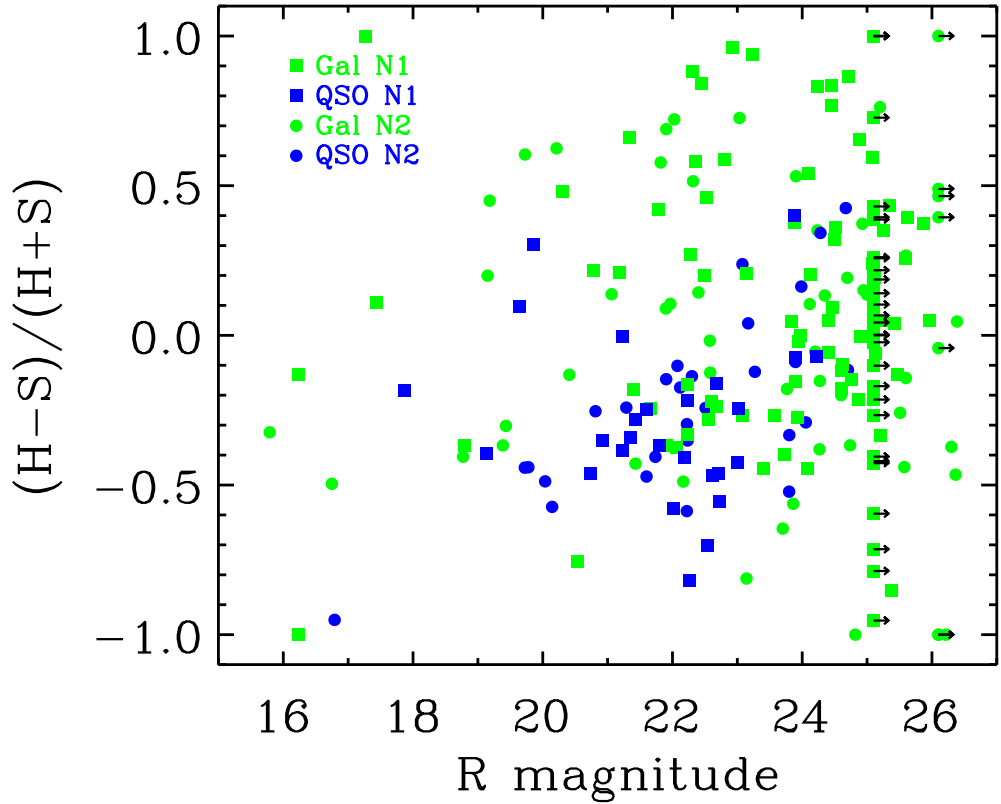} 

{\caption {Hardness ratio against optical magnitude for ELAIS Chandra
sources.}}

\end{figure}

An analysis of the hardness ratios, $HR=[(H-S)/(H+S)]$, of the Chandra
sources shows that $HR$ increases at fainter fluxes (as seen in the
CDFS -- Tozzi et al. 2001). The simplest explanation of this effect is
that sources get harder at lower luminosities, since the redshift
distribution is unlikely to change significantly from $10^{-14}~ {\rm
to}~ 10^{-15}~ {\rm erg ~cm}^{-2}~{\rm s}^{-1}$. It is worth pointing
out that $HR$ alone is not a good measure of the absorption of an
absorbed power-law X-ray source. Due to the negative k-correction of a
highly absorbed source, sources at high redshift will have much lower
values of $HR$ than sources with the same rest-frame spectra at lower
redshifts. In Figure 3 we plot the hardness ratio $HR$ against the $R$
magnitude. Note that the hardest sources are galaxies, indicating a
lack of quasars with hard spectra. Many of the hard ($HR>0$) galaxies
are very faint, which suggests they are at high redshifts ($z>1$) and
have large absorbing columns. There is no overall correlation of $HR$
with magnitude for the galaxies, although this non-correlation is not
trivial to interpret given the positive correlation between $R$ and
$z$ for galaxies and the negative correlation between $HR$ and $z$ for
a given absorbing column. This could be indicating a more rapid
evolution of hard sources than soft sources, as required by some XRB
models (e.g. Gilli et al. 2001).

The hardness ratio of an unabsorbed, broad line quasar is typically
$HR=-0.5$. Most of the optically bright quasars ($R<23$) in Fig. 3
show values similar to this. However, the fainter quasars tend to have
harder X-ray spectra such that at $R=24$, $HR\approx 0$. This cannot
be a redshift effect because $HR$ decreases at high redshift and if
these sources were at low redshift then their galaxies would outshine
the nucleus in the optical. A simple explanation for this change is
that some of the optically fainter quasars are undergoing absorption
in their environments, both in the optical by dust and in the X-rays
by cool gas. The nuclear region in such sources may suffer a few
magnitudes of extinction or may be totally obscured, in which case the
light we observe is either scattered into our line-of-sight or emitted
from a region beyond the obscuring medium. For a galactic gas-to-dust
ratio, the large absorbing columns necessary for such hard X-ray
spectra would equate to very high optical reddening, suggesting that
either the light we observe is scattered or there is a very high
gas-to-dust ratio as found in some low-redshift Seyferts (Maiolino et
al. 2001).

\section{The fraction of high-luminosity Type 2 quasars}

Models of the X-ray background have long postulated the existence of
high-luminosity ($L_{2-10{\rm keV}}>10^{44} {\rm erg~cm}^{-2}~{\rm
s}^{-1}$), obscured Type 2 quasars, showing only narrow emission
lines. Within the unified scheme for AGN, it is expected that the
central regions of Type 2 quasars are obscured along our line-of-sight
by a dusty torus (e.g. Antonucci 1993). However, previous hard X-ray
surveys with ASCA and BeppoSax have revealed few such objects leading
some authors to doubt their existence at all (Halpern et al. 1999).

Of course, radio-loud Type 2 quasars (i.e. radio galaxies) have been
known for a long time and it is worth considering the known properties
of high-redshift radio galaxies when searching for similar objects
which are radio-quiet. The ratio of narrow-line to broad-line
high-luminosity radio-loud AGN is well determined by virtue of the
isotropy of extended low-frequency radio emission and is 1.5 (Barthel
1989). At lower radio luminosities (151MHz luminosities $L_{151}<3
\times 10^{26}$ W Hz$^{-1}$ sr$^{-1}$) this ratio increases to
$\approx 7$ (Willott et al. 2000). There is no evidence for evolution
of this ratio as a function of redshift.

The perceived picture of the optical appearance of high redshift radio
galaxies is biased towards the most luminous objects (e.g. 3C radio
galaxies) which have been most widely studied. High-redshift 3C radio
galaxies have very strong high-ionization emission lines and tend to
have blue optical-to-near-infrared colours due to extended rest-frame
UV emission which is either non-stellar in origin or due to
jet-triggered star-formation (e.g Best et al. 1998). Due to the
correlation between emission line strengths and radio luminosity
(Rawlings \& Saunders 1991), lower luminosity radio galaxies have much
weaker lines. They also have less rest-frame UV emission and a large
fraction of them are classified as extremely red galaxies ($R-K>5$)
with colours typical of old galaxies at $z>1$ (Willott et al. 2001).

\begin{figure} 
\hspace{0.5cm} 
\vspace{-0.1cm} 
\epsfxsize=0.9\textwidth
\epsfbox{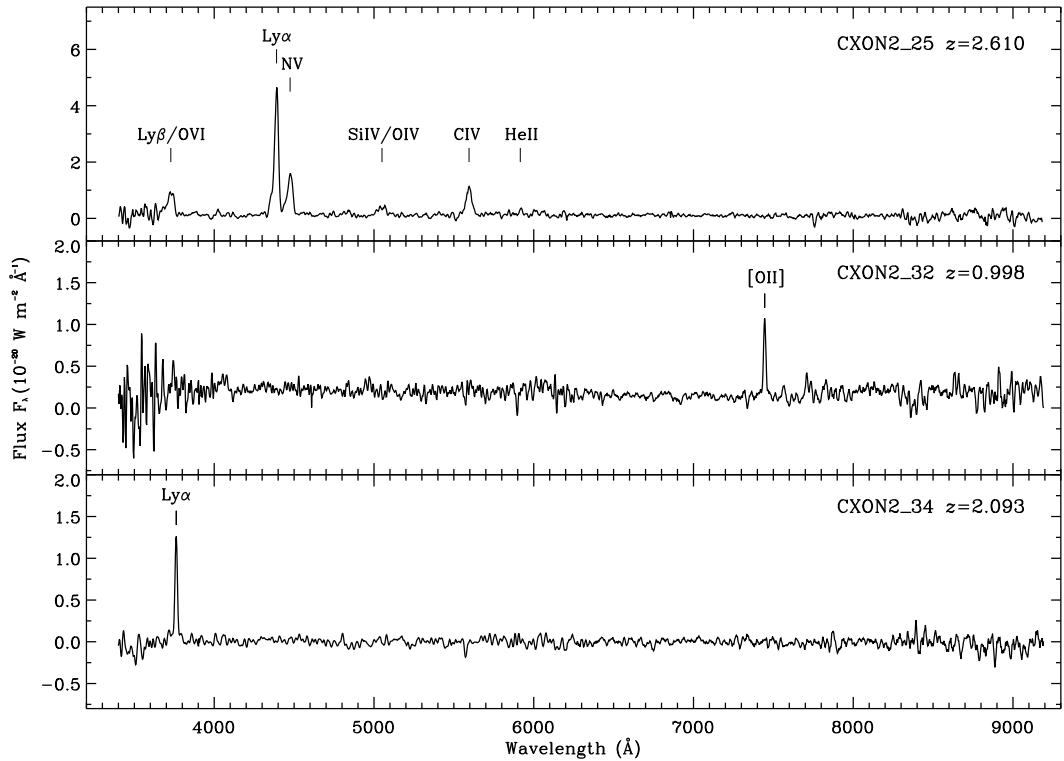} 
\vspace{1.3cm} 

{\caption {Optical spectra of hard X-ray selected ELAIS Chandra
sources.  The sources are radio-quiet Type 2 quasars. }}

\end{figure}

The identification of hard Chandra sources with galaxies which are
faint in the optical and very red (Giacconi et al. 2001; Crawford et
al. 2000; Barger et al. 2001) suggests that many of these may also be
evolved galaxies at $z>1$. Cowie et al. (2001) present high quality
data on a couple of such objects showing that they do indeed appear to
be red galaxies with little contribution to the broad band magnitudes
by AGN emission. Although such objects do not always show strong
narrow emission lines, their high redshifts imply high X-ray
luminosities for samples with flux-limits similar to the ELAIS Chandra
observations. Accounting for the hard X-ray absorption, (which can be
a factor of 10 for Type 2 quasars detected at 2-10 keV; Norman et
al. 2001) then the intrinsic luminosities would be very high and well
above the $L_{2-10{\rm keV}}>10^{44} {\rm erg~cm}^{-2}~{\rm s}^{-1}$
criterion. The large number of faint galaxy counterparts in Fig. 2
suggest that a sizeable fraction of ELAIS Chandra sources will be Type
2 quasars at $z>1$.

We have obtained optical spectra from the WHT for a small sample of
five hard X-ray selected ELAIS Chandra sources. Redshifts were
determined for three of the five sources (the two undetected sources
have $R \sim25$ and very red optical to near-IR colours). Spectra of
the three sources with emission lines are shown in Figure 4. Two
sources have spectra which are very similar to those of high-redshift
radio galaxies, with narrow (FWHM $< 1500$ kms$^{-1}$)
lines. CXON2\_25 has slightly broader emission lines ($1800-2500$
kms$^{-1}$) and is most likely a heavily reddened quasar. The X-ray
luminosities of the two $z>2$ X-ray sources are $L_{2-10{\rm
keV}}>10^{44} {\rm erg~cm}^{-2}~{\rm s}^{-1}$ and that of the $z=1$
source is $>10^{43} {\rm erg~cm}^{-2}~{\rm s}^{-1}$. Accounting for
the large absorption depression of the hard X-ray luminosity (all have
$HR>0$), all three sources are classified as Type 2 quasars. We expect
that complete spectroscopic follow-up of the survey will lead to the
discovery of many more such objects and determine their redshift
distribution. Folding in the knowledge of the amount of absorption
from the high quality X-ray spectra which will come from our deep
XMM-Newton pointing, we will be able to derive the relative space
densities of Type 2 to Type 1 sources and its dependence upon redshift
and luminosity.

\section{Studies of Type 2 quasar hosts}

\begin{figure}[t]
\vspace{-0.1cm} 
\epsfxsize=1.0\textwidth
\epsfbox{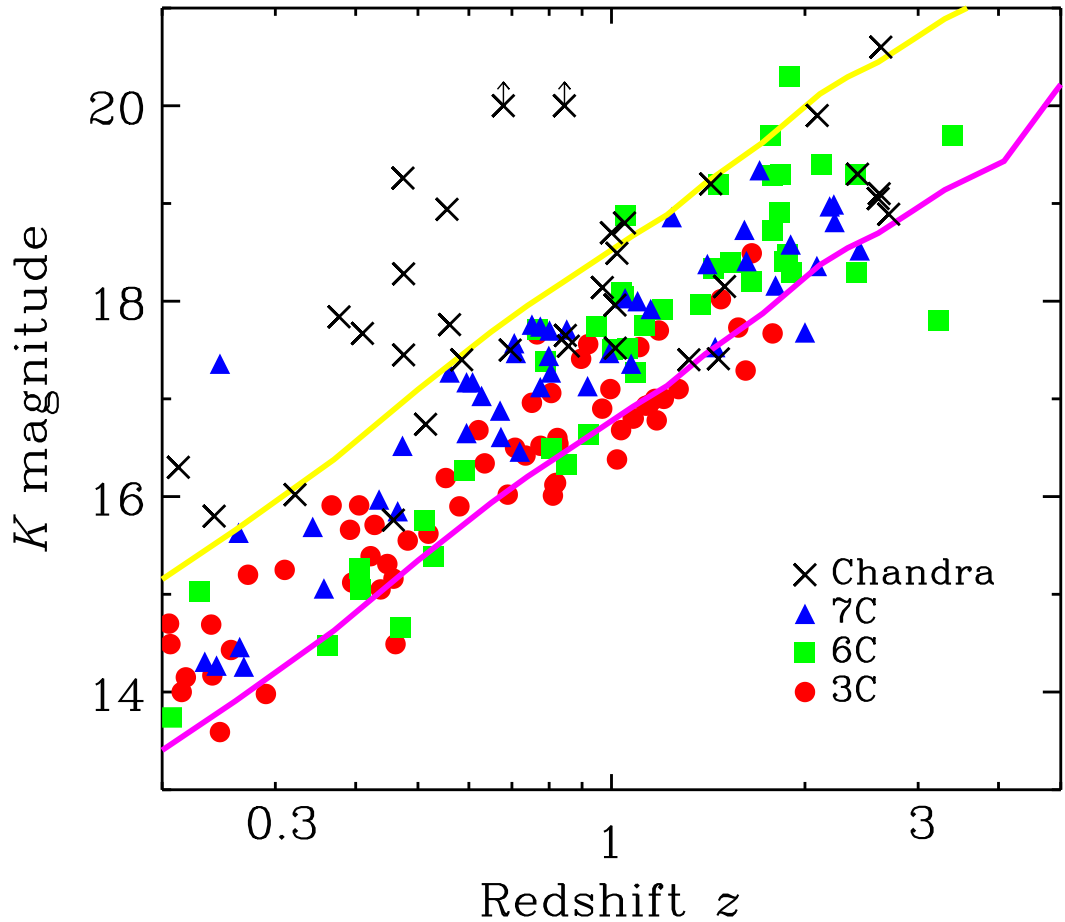} 

{\caption {The near-infrared Hubble diagram for radio galaxies from
the 3C, 6C and 7C samples and Type 2 X-ray sources from Chandra
surveys. Note that the X-ray source host galaxies have a much larger
range of $K$-band luminosities than the radio galaxies. The curves are
passive evolution models for galaxies which formed all their stars in
a short burst at high redshift ($z=10$). The upper curve corresponds to 
$L_{\star}$ and the lower curve to $5L_{\star}$}}

\end{figure}

The large numbers of Type 2 quasars which will be identified in deep
hard X-ray surveys provide a new opportunity to study the host
galaxies of AGN, without the difficulties of PSF subtraction which
complicate studies of Type 1 quasar hosts. The near-infrared Hubble
diagram (or $K-z$ relation) of radio galaxies up to $z \sim 4$ is
consistent with passive evolution of massive galaxies with
luminosities in the range $2 - 5L_{\star}$ (Jarvis et al. 2001). By
comparing 3C radio galaxies with fainter samples (6CE, 7CRS) there is
evidence for a weak correlation of stellar $K$-band luminosity with
radio luminosity (Eales et al. 1997; Willott et al., in prep.). This
correlation is likely to have its origins in the black hole mass --
bulge mass -- stellar velocity dispersion correlations (Magorrian et
al. 1998; Gebhardt et al. 2000) since the $K$-band luminosity is a
good indicator of the stellar mass. 

In Figure 5 we show the near-infrared Hubble diagram for an incomplete
sample of Chandra sources without broad emission lines (ELAIS Deep;
Hornschemeier et al. 2001; Barger et al. 2001; Cowie et al. 2001). The
X-ray sources have a wider range of $K$-band luminosities than the
radio galaxies, from sub-$L_{\star}$ to 5$L_{\star}$. This can be
explained in terms of the correlation found for radio galaxies
mentioned above. The X-ray sources range from very low luminosity to
powerful AGN and it is likely that powerful AGN only occur in the more
massive galaxies which have large black holes. The sub-$L_{\star}$
host galaxies are generally at low redshifts, where the correlations
between luminosity and redshift in the samples mean that they have low
X-ray luminosities. However, it is worth remembering that these
samples are nowhere near complete in terms of spectroscopic redshifts
and obtaining redshifts for low-luminosity X-ray sources at $z>1$ will
be very difficult. Confirmation of this result will be possible when
complete samples are available and emission line contributions to the
magnitudes at $K$-band can be estimated.


\end{document}